\documentclass[12pt]{article}
\usepackage[]{graphicx}
\usepackage{amssymb}
\usepackage{amsmath}
\title{Investigation of the spectra of coupled polaritons on the periodically modulated %
metallic layer and the narrow regions of anomalous transparency}

\author{A. V. Kats, Y. V. Bludov}
\date{}
\begin{document}

\maketitle

\begin{center} \small{
Usikov Institute for Radiophysics and Electronics National Academy of Sciences of Ukraine \\ 12 Ak. Proskury Street,
61085, Kharkiv, Ukraine. E-mail: bludov@ire.kharkov.ua}
\end{center}

\begin{abstract}
The paper deals with the theoretical investigation of plane,
normally incident electromagnetic wave transmission through the
flat metal film whose dielectric constant has small periodical
sinusoidal modulation in one dimension parallel to the projection
of the electric field onto the film surface. The dependencies of
the film transmittancy on the parameters of the problem
(frequency, modulation depth and absorption) are examined. It is
shown that the film transmittancy increases considerably when the
conditions for resonance interaction of an incident
electromagnetic wave with surface plasmon polaritons (SPPs) are
met. It is found that for small but finite absorption there are
two frequencies in the vicinity of which the transmittancy can
achieve the values of the order of unity due to resonances on
symmetric and antisymmetric (relative to the mean plane) SPP
modes. It is shown that for each value of absorption there exists
a certain optimal modulation depth, which maximizes the resonance
transparency.
\end{abstract}


\section{Introduction}

A metal-dielectric interface is known to be able to sustain
surface plasmon polaritons (SPPs) \cite{ritchie}, i.e., collective
excitations of surface electron density coupled to the
electromagnetic field. Since the SPP phase and group velocities
are less than the velocity of light, the SPPs cannot be excited by
external electromagnetic wave falling on a perfectly flat metal
surface \cite{maradudin,rather}. Nevertheless, when the film is
periodically modulated and a spatial period of modulation
coincides with the wavelength of SPP, the latter can be excited by
a normal incident electromagnetic wave. This leads to the
well-known Wood anomalies discovered in 1902 for the reflectance
spectra of metallic gratings illuminated by $p$- ($TM$-) polarized
light \cite{wood}. Later these anomalies were associated with the
SPP excitation \cite{fano}. Since then both reflection
\cite{hessel,watts,spa_exp,ist_spis} and transmission
\cite{ist_spis,schroter,treacy,porto,collin,went} gratings are of
great interest to physicists. In reflection gratings two types of
anomalies have been predicted theoretically
\cite{hessel,watts,ist_spis} and later confirmed experimentally
\cite{spa_exp}: the effect of SPP excitation is responsible for
the existence of one anomaly\footnote{Here it is necessary to
underline that except the SPP excitation there exists another type
anomaly related to the parameters (wavelength, angle of incidence)
such that one of the diffracted orders intersects the boundary
between propagating and evanescent waves. From the formal point of
view it corresponds to the branch point in a Fourier expansion and
was first indicated by Lord Rayleigh in discussion on Wood's
experiments. For good metals this anomaly is close to that related
to the SPP excitation. Note that for infinitely conducting media
these two anomalies coincide. These two anomalies are often called
as Wood anomaly (in a broad sense); it seems better to call the
former Wood anomaly and the latter Rayleigh anomaly.}, and the
other one exists when the depth of the grating grooves exceeds
some critical value (of order of the quarter wavelength). The
latter anomaly is associated with the standing waves having the
field energy highly concentrated inside the grooves. Transmission
gratings exhibit an unusual property: this kind of structures can
be highly transmitting at certain frequencies
\cite{schroter,treacy}. As it was originally shown in paper
\cite{porto}, there are two types of transmission resonances in
the transmission grating: coupled SPP's on both horizontal
surfaces of the metallic grating, and cavity modes located inside
the slits. The authors of paper \cite{collin} preferred to
identify these resonances as horizontal and vertical SPP
resonances, respectively. Recently these resonances have been
confirmed experimentally \cite{went}.

Another kind of transmitting structures was proposed recently: a
flat, optically thick metal film, perforated with the
subwavelength hole 2D array
\cite{ebbesen,ghaemi,kim,krishnan,hohng}. The transmission of
light through these structures (due to SPP resonance) can be much
larger than that expected from the standard aperture theory
\cite{bethe}. Notice that this extraordinary transmission effect
has potential applications \cite{sambles,kitson,thio} in
subwavelength photolithography, near-field microscopy,
wavelength-tunable filters, optical modulators, and flat-panel
displays.

To obtain the high transmittancy of the metal film, instead of
hole array one can use the films with spatially modulated
dielectric constant \cite{shubin,zay}. The transmission of the
electromagnetic wave through the flat metal film, whose dielectric
constant has 1D modulation of sinusoidal shape, was considered in
paper \cite{shubin}. It has been found that at the SPP resonance
such dissipation-free films become fully transparent regardless of
its thickness; however, the width of transmission resonance
shrinks when the film thickness increases. Recently, an attempt of
analytical treatment of optical transmission through periodically
modulated metal films capable of supporting SPPs has been
presented in paper \cite{zay}. However, in the above-mentioned
paper the coupling between SPPs on different film boundaries has
not been taken into account explicitly. Also the authors of papers
\cite{shubin,zay} neglected the field dissipation in the metal
film and obtained the expression for film transmittancy without
taking into consideration the off-resonance second spatial field
harmonic.

In this paper we examine theoretically  transmission of a plane
normally incident electromagnetic wave through the flat metal film
with dielectric constant being periodically modulated in one
in-plane dimension. We analyze the transmittance in the case where
the periodicity is caused by a sinusoidal shape modulation. The
dependencies of the film transmittancy upon incident wave
frequency, modulation depth and film thickness are investigated.
It is found that the film transmittancy increases considerably
when the conditions for resonance interaction between the incident
electromagnetic wave and SPPs are satisfied and light absorbtion
is small. It is shown that the dependence of the film
transmittancy upon incident electromagnetic wave frequency
exhibits two maxima, which correspond to the resonances on
symmetric and antisymmetric SPP modes. The values of the film
transmittancy in the vicinity of maxima can be much higher than
that of the unmodulated film. It is shown that when the losses in
the metal are taken into account, a certain optimal modulation
depth (which maximizes the film resonance transparency)
corresponds to each value of the losses.

The paper is organized as follows. In section 2 we derive the
algebraic expression for the modulated film transmittancy. In
section 3 the transmission properties of the film are analyzed and
the numerical results are presented. The paper is concluded with a
brief summary of results and possible applications (section 4).

\section{Problem statement and main equations}

Consider a flat metallic film of a finite thickness $d$, supposing
that the film/air boundaries coincide with the planes $z=0$ and
$z=d$ (see Fig.1). Besides, the dielectric constant of the film is
considered to be modulated along direction $x$ as
\begin{equation}
\varepsilon(x)=\varepsilon_m^{\prime}(1-i\delta+a\cos(gx)).
\end{equation}
Here $\varepsilon_m=\varepsilon_m^{\prime}(1-i\delta)$ is the
complex dielectric permittivity of an unmodulated metal
($\varepsilon_m^{\prime}<0$), $\delta$ is the loss angle tangent,
$a$ is the modulation depth, $g=2\pi/b$ is the modulation
wavenumber ($b$ is the period of modulation).

The plane p-polarized electromagnetic wave is assumed to propagate
in a positive $z$-axis direction. Since the solution of our
problem has to be uniform along direction $y$, we can assign
$\frac{\partial}{\partial y}\equiv 0$. Besides, we suppose that
$\vec{E},\vec{H} \sim \exp(-i\omega t)$. In this case the Maxwell
equations for p-polarized electromagnetic wave can be represented
in the form ($E_{y} = H_{x} = H_{z} = 0$):

\begin{eqnarray}
& \displaystyle{\frac{\partial H_y(x,z)}{\partial
z}=\frac{i\omega}{c}\varepsilon(x)E_x(x,z),} & \label{eq:m-ras1}\\
& \displaystyle{\frac{\partial H_y(x,z)}{\partial
x}=-\frac{i\omega}{c}\varepsilon(x)E_z(x,z),} &
\label{eq:m-ras2}\\ & \displaystyle{\frac{\partial
E_z(x,z)}{\partial x}-\frac{\partial E_x(x,z)}{\partial
z}=-\frac{i\omega}{c}H_y(x,z).} \label{eq:m-ras3} &
\end{eqnarray}

Since the dielectric permittivity  is  periodic,  the solution of
equations (\ref{eq:m-ras1})--(\ref{eq:m-ras3}) can be given in the
form of Fourier--Floquet series. Due to supposed smallness of the
modulation, we can truncate this expansion taking into account
three spatial field harmonics only. Thus, the $y$-component of the
magnetic field can be written as
\begin{equation}\label{eq:hy-fou}
H_y(x,z)=H_0(z) + H_1(z)\cos(gx) + H_2(z)\cos(2gx).
\end{equation}
Here $H_0(z)$, $H_1(z)$,$H_2(z)$ are the amplitudes of zeroth,
first and second spatial harmonics of magnetic field,
respectively.

Substituting the expression (\ref{eq:hy-fou}) in the Maxwell
equations, and neglecting the terms of $a^2$ order, we obtain
\begin{eqnarray}
&\displaystyle{H_0^{\prime\prime}(z)+k^2\varepsilon_mH_0(z)-\alpha
H_1^{\prime\prime}(z)=0,}& \label{eq:har0}\\
&\displaystyle{H_1^{\prime\prime}(z)-\left(g^2-k^2\varepsilon_m\right)H_1(z)-2\alpha
H_0^{\prime\prime}(z)-\alpha \left[H_2^{\prime\prime}(z)
-2g^2H_2(z)\right]=0,}& \label{eq:har1}\\
&\displaystyle{H_2^{\prime\prime}(z)-\left(4g^2-k^2\varepsilon_m\right)H_2(z)-\alpha\left[H_1^{\prime\prime}(z)-2g^2H_1(z)\right]=0.}&
\label{eq:har2}
\end{eqnarray}
Here $\alpha=(a/2)(1-i\delta)^{-1}$, and the primes denote the
derivatives with respect to $z$.

The eigenvalues of the homogeneous set of equations
(\ref{eq:har0})-(\ref{eq:har2}) can be found from the bicubic
characteristic equation
\begin{equation}
\left(\gamma^2-P_0^2\right)\left(\gamma^2-P_1^2\right)\left(\gamma^2-P_2^2\right)-%
\alpha^2\left[\left(\gamma^2-P_0^2\right)\left(\gamma^2-2g^2\right)^2+%
2\gamma^4\left(\gamma^2-P_2^2\right)\right]=0, \label{che}
\end{equation}
where
    $$
P_n^2=n^2g^2-k^2\varepsilon_m,\qquad n=0,1,2.
    $$
The roots of the characteristic equation (\ref{che}) can be written to within $\alpha^2$ as
    $$
\gamma_0^2= P_0^2-2\alpha^2\frac{P_0^4}{g^2}+O(\alpha^4), \quad \gamma_1^2=
P_1^2-\alpha^2\frac{[P_1^2-2g^2]^2}{3g^2}+2\alpha^2\frac{P_1^4}{g^2}+O(\alpha^4),
    $$
    $$
\gamma_2^2= P_2^2+\alpha^2\frac{[P_2^2-2g^2]^2}{3g^2}+O(\alpha^4).
    $$
The general solution of equations (\ref{eq:har0})-(\ref{eq:har2})
can be given in the form (since $|\alpha| \ll 1$, for our purposes
it is sufficient to restrict our consideration of the solution to
the linear terms in $\alpha$)
\begin{equation}
H_0(z)=\left\{B^{(s)}_0\sinh{\gamma_0z}+B^{(c)}_0\cosh{\gamma_0z}\right\}+%
\alpha\phi_{01}\left\{B^{(s)}_1\sinh{\gamma_1z}+B^{(c)}_1\cosh{\gamma_1z}\right\},
\end{equation}
\begin{eqnarray}
H_1(z)&=&2\alpha\phi_{10}\left\{B^{(s)}_0\sinh{\gamma_0z}+B^{(c)}_0\cosh{\gamma_0z}\right\}+%
\left\{B^{(s)}_1\sinh{\gamma_1z}+B^{(c)}_1\cosh{\gamma_1z}\right\}+\nonumber\\%
&&\alpha\phi_{12}\left\{B^{(s)}_2\sinh{\gamma_2z}+B^{(c)}_2\cosh{\gamma_2z}\right\},
\end{eqnarray}
\begin{equation}
H_2(z)=\alpha\phi_{21}\left\{B^{(s)}_1\sinh{\gamma_1z}+B^{(c)}_1\cosh{\gamma_1z}\right\}+%
\left\{B^{(s)}_2\sinh{\gamma_2z}+B^{(c)}_2\cosh{\gamma_2z}\right\}.
\end{equation}
Here
    $$
\phi_{nl}=\frac{\gamma_l^2-nlg^2}{\gamma_l^2-P_n^2},\qquad n,l=0,1,2,
    $$
$B^{(s)}_n$, $B^{(c)}_n$ are the arbitrary coefficients
corresponding to the eigenfunctions of equations
(\ref{eq:har0})-(\ref{eq:har2}). The coefficients $B^{(s)}_n$,
$B^{(c)}_n$ are to be obtained by matching the boundary
conditions.

The fields outside the film can be found from equations
(\ref{eq:har0})--(\ref{eq:har2}), if we put $\varepsilon(x)=1$ and
$\alpha=0$. Apart from this, for the first and the second spatial
harmonics we have to use radiation boundary conditions. So, the
spatial harmonics of magnetic field outside the film are
\begin{eqnarray}
&\displaystyle{H^-_0(z)=H_i\exp\left[-p_0z\right]+H_r\exp\left[p_0z\right],
\quad H^-_1(z)=H^-_1\exp\left[p_1z\right],}&
\\
&\displaystyle{H^-_2(z)=H^-_2\exp\left[p_2z\right], \qquad z \le
0,}& \nonumber
\end{eqnarray}
\begin{eqnarray}
&\displaystyle{H^+_0(z)=H_t\exp\left[-p_0\left(z-d\right)\right],
\quad H^+_1(z)=H^+_1\exp\left[-p_1\left(z-d\right)\right],}&
\\
&\displaystyle{H^+_2(z)=H^+_2\exp\left[-p_2\left(z-d\right)\right],
\qquad z \ge d.}& \nonumber
\end{eqnarray}
Here $H_i$, $H_r$, $H_t$ are the amplitudes of incident,
transmitted and reflected waves, correspondingly, $H_1^{\pm}$,
$H_2^{\pm}$ are the amplitudes of the first and the second
harmonics, $p_n=\sqrt{n^2g^2-k^2}$, $\rm{Re}(p_n)-\rm{Im}(p_n)\geq
0$, $n=0,1,2$. The tangential component of electric field can be
obtained from the equation (\ref{eq:m-ras1}) and with accuracy up
to the linear in $\alpha$ terms can be represented in the form
\begin{equation}
E_x=\frac{1}{ik\varepsilon_m}\left[H_0^{\prime}(z)-\alpha H^{\prime}_1(z)+\cos(gx)\left\{H^{\prime}_1(z)-2\alpha H_0^{\prime}(z)%
-\alpha H^{\prime}_2(z)\right\}+\cos(2gx)\left\{H^{\prime}_2(z)-%
\alpha H^{\prime}_1(z)\right\}\right].
\end{equation}
The continuity of the tangential components of the electric and
magnetic field at boundaries $z=0$ and $z=d$ results in the
following system of linear equations
\begin{equation}
\widehat{H}_0X_0+\alpha\widehat{A}_{01}X_1=G. \label{eq:mat0}
\end{equation}
\begin{equation}
2\alpha\widehat{A}_{10}X_0+\widehat{H}_1X_1+\alpha\widehat{A}_{12}X_2=0,
\label{eq:mat1}
\end{equation}
\begin{equation}
\alpha\widehat{A}_{21}X_1+\widehat{H}_2X_2=0. \label{eq:mat2}
\end{equation}
Here \setlength{\arraycolsep}{5pt}$$ \widehat{H}_l=\left(
\begin{array}{cccc}
1&0&-1&0\\ 0&\eta_l&p_l&0\\
\displaystyle{\cosh{\gamma_ld}}&\displaystyle{\sinh{\gamma_ld}}&0&-1\\
\displaystyle{\eta_l\sinh{\gamma_ld}}&\displaystyle{\eta_l\cosh{\gamma_ld}}&0&-p_l\\
\end{array} \right), \quad \widehat{A}_{nl}=\left(
\begin{array}{cccc}
\phi_{nl}&0&0&0\\ 0&\eta_l\Phi_{nl}&0&0\\
\displaystyle{\phi_{nl}\cosh{\gamma_ld}}&\displaystyle{\phi_{nl}\sinh{\gamma_ld}}&0&0\\
\displaystyle{\eta_l\Phi_{nl}\sinh{\gamma_ld}}&\displaystyle{\eta_l\Phi_{nl}\cosh{\gamma_ld}}&0&0\\
\end{array} \right),$$
$$X_0=\left(\begin{array}{c} B^{(c)}_0 \\ B^{(s)}_0 \\ H_r \\ H_t
\end{array}\right), \quad X_1=\left(\begin{array}{c} B^{(c)}_1 \\ B^{(s)}_1 \\ H^-_1 \\ H^+_1
\end{array}\right), \quad X_2=\left(\begin{array}{c} B^{(c)}_2 \\ B^{(s)}_2 \\ H^-_2 \\
H^+_2
\end{array}\right), \quad  G=\left(\begin{array}{c}
H_i \\ p_0H_i \\ 0 \\ 0
\end{array}\right),$$ $$\Phi_{nl}=\phi_{nl}-1,\quad \eta_n=-\gamma_n/\varepsilon_m$$
The solution of equations (\ref{eq:mat0})--(\ref{eq:mat2}) for the
fundamental (zeroth), the first and the second spatial harmonics
can be written as
\begin{equation}
X_0=\left[\widehat{I}+2\alpha^2\widehat{H}_0^{-1}\widehat{A}_{01}\widehat{F}^{-1}%
\widehat{A}_{10}\right]\widehat{H}_0^{-1}G, \label{mat-sol0}
\end{equation}
\begin{equation}
X_1=-2\alpha\widehat{F}^{-1}%
\widehat{A}_{10}\widehat{H}_0^{-1}G, \label{mat-sol1}
\end{equation}
\begin{equation}
X_2=2\alpha^2\widehat{H}_2^{-1}\widehat{A}_{21}\widehat{F}^{-1}%
\widehat{A}_{10}\widehat{H}_0^{-1}G, \label{mat-sol2}
\end{equation}
where
\begin{equation}
\widehat{F}=\widehat{H}_1-2\alpha^2\widehat{A}_{10}
\widehat{H}_0^{-1}\widehat{A}_{01}-\alpha^2\widehat{A}_{12}%
\widehat{H}_2^{-1}\widehat{A}_{21},
\end{equation}
$\widehat{I}$ is the unit $4 \times 4$ matrix. The main features
of electromagnetic wave transmission through the modulated film
can be deduced from the analysis of expression (\ref{mat-sol0}).
Matrices $\widehat{H}_0$ and $\widehat{H}_2$ are the nonsingular
matrices. At the same time matrix $\widehat{H}_1$ is a singular
matrix. The condition $\rm{Det}|\widehat{H}_1|=0$,
$\left[p_1\tanh{\gamma_1d}-\eta_1\right]\left[p_1-\eta_1\tanh{\gamma_1d}\right]=0$
is known to be the dispersion relation for SPP (with wavenumber
$g$ and frequency $\omega$) on the unmodulated film \cite{cam}.
Thus, if the frequency of an incident wave is far enough from the
SPP frequency (i.e. the value of $\rm{Det}|\widehat{F}| $ is far
from zero), the term, proportional to $\alpha^2$ in expression
(\ref{mat-sol0}) can be neglected and then $X_0 \approx
\widehat{H}_0^{-1}G$. In other words, the transmittancy of the
film with modulated dielectric permittivity is approximately equal
to that of the unmodulated film. If  the incident wave frequency
is close to that of the SPP symmetric or antisymmetric mode of the
modulated film (i.e. $\rm{Det}|\widehat{F}| \approx 0$) then the
term proportional to $\alpha^2$ achieves the values of order of
unity and becomes comparable with the first term. Therefore, the
transmittancy of the film with modulated dielectric permittivity
differs considerably from that of the unmodulated film.

\section{Results}

In this section we investigate the dependence of film transmittancy upon frequency, modulation depth, film thickness
and losses.

\textbf{The dependence of the modulated film transmittancy upon
the frequency of incident wave} is presented in Fig.2. For the
comparison the frequency dependence of the unmodulated film
transmittancy for the same values of parameters is depicted in
Fig.2 by dashed lines. As evident from Fig.2, the frequency
dependence of modulated film transmittancy exhibits two resonance
maxima. At the same time the value of the modulated film
transmittancy at frequencies close to the resonance frequencies is
much higher than the value of uniform unmodulated film
transmittancy. In particular, when there are no losses in the film
(curve 1), the value of the modulated film transmittancy at
resonances is equal to unity. In other words, the full
transmission of incident electromagnetic wave through the lossless
film with modulated dielectric permittivity is possible at two
fixed frequencies.  The reason for full transparency of the metal
film with modulated dielectric permittivity lies in the following.
The SPP spectrum of a finite thickness film contains two modes,
the high-frequency (HF) symmetric mode and the low-frequency (LF)
antisymmetric mode \cite{cam}. The SPP symmetric mode  is
characterized by the fact that the values of the SPP magnetic
field at film boundaries $z=0$ and $z=d$ are equal, and
$z$-components of the electric field (along with the surface
charge density) are of different signs, i.e., the HF eigenmode
possesses reflective symmetry with respect to the plane $z = d/2$.
At the same time the values of the SPP antisymmetric mode magnetic
field at the film boundaries have opposite signs. So, the LF
eigenmode possesses reflective antisymmetry with respect to the
plane $z=d/2$. Nevertheless, the absolute values of the SPP
electric and magnetic field amplitudes at the film boundaries
$z=0$ and $z=d$ are equal to each other both for the SPP symmetric
mode and for the SPP antisymmetric one. Hence, in the case where
the film modulation period is equal to the SPP wavelength (i.e.
the conditions for the SPP excitation are met) the incident wave
upon diffracting on the modulation excites the SPP at boundary
$z=0$. Further, following the interaction with the modulation, at
boundary $z=d$ the SPP transforms into the bulk electromagnetic
wave, this electromagnetic wave being radiated with the same
amplitude as an incident wave (due to the above-mentioned equality
between the SPP field amplitudes at both of the boundaries). Thus,
the transmittancy of the lossless film with modulated dielectric
permittivity is equal to unity if the absorbance in the film is
absent and when the conditions for the SPP excitation are met.
Fig.2 also shows that as the losses increase, the modulated film
transmittancy decreases monotonically. At the same time when
losses in the film are taken into account, the maximum value of
transmittancy for the LF resonance exceeds that for the HF
resonance.

The necessity to allow for the second spatial harmonic in the
expression (\ref{eq:hy-fou}) is seen from Fig.3, which depicts the
frequency dependence of the second spatial harmonic amplitude at
boundary $z=0$ in the case where the losses in the film do not
exist. As seen from Fig.3, the amplitude of the second spatial
harmonic at frequencies far from the SPP resonance frequencies are
considerably less than the amplitude of the incident wave. At the
same time the value of the second spatial harmonic amplitude in
the vicinity of resonances is of the same order as the amplitude
of the incident wave. The need to take into account the second
spatial harmonic, when the expressions for the modulated film
transmittancy and reflectance are obtained, follows from this
ratio between the amplitudes of the second harmonic and the
incident wave.

\textbf{The influence of the modulation depth value upon the modulated film transmittancy and upon the value of
resonance peak frequency} is shown in Fig.4. As Fig.4 illustrates, as the modulation depth $a$ increases, the resonance
peak frequency decreases monotonically both in the case of LF resonance (curves 1$'$,2$'$ in Fig.4a) and in the case of
HF resonance (curves 3$'$,4$'$ in Fig.4b). As it follows from the comparison of curves 1$'$,2$'$, the LF resonance peak
frequency is increased, as the film thickness increases (Fig.4a). At the same time the HF resonance peak frequency is
decreased, as the film thickness increases (Fig.4b). Note that at LF resonance (Fig.4a), as the modulation depth $a$
increases, the value of the transmittance maximum $T_{max}$ increases as well (at small values of $a$). However,
starting from some value of the modulation depth $a_{opt}$ ($a_{opt} \approx 0.072$ for $\omega_0d/c=0.5$ and $a_{opt}
\approx 0.073$ for $\omega_0d/c=1.0$) the value of $T_{max}$ decreases with an increase of $a$. At HF resonance
(Fig.4b), the value of $T_{max}$ reaches its maximum, when $a_{opt} \approx 0.076$ (for $\omega_0d/c=0.5$) and when
$a_{opt} \approx 0.072$ (for $\omega_0d/c=1.0$). Thus, both for LF resonance and for HF resonance with the losses in
the film being taken into account, there exists a certain value $a_{opt}$, which corresponds to the ``optimal tuning":
when the modulation depth of the film is equal to $a_{opt}$, the resonance transmission of the electromagnetic wave
through the modulated film is highest for the given value of losses. This phenomenon can be explained as follows: when
the film modulation depth is equal to $a_{opt}$, the SPP excitation occurs with the highest effectiveness due to the
equality of free space impedance and modulated film impedance. Besides, Fig.4a shows that the value of $T_{max}$ at LF
resonance is decreased, as the film thickness $d$ increases. At the same time in the case of the HF resonance, as the
film thickness increases, the value of $T_{max}$ increases likewise (Fig.4b).

This phenomenon is illustrated in detail in Fig.5, which presents the dependence of transmittancy maximum value and
resonance peak frequency upon the film thickness. As seen from Fig.5, with an increase of the film thickness, the value
of transmittance maximum $T_{max}$ is decreased monotonically for the LF resonance (curve 1). At the same time at the
HF resonance the value of $T_{max}$ is monotonically increased with an increase of $d$ (curve 2) and in the limit $d
\to \infty$ the values of $T_{max}$ for two resonances coincide. With an increase in $d$ the resonance peak frequency
increases in the case of the LF resonance (curve 1$'$) and decreases in the case of the HF resonance (curve 2$'$).

\textbf{The dependence of the optimal modulation depth upon the losses} is depicted in Fig.6 in the cases of LF and HF
resonances. As Fig.6 shows, at small values of losses the value of optimal modulation depth does not depend essentially
upon the value of losses. However, at high values of losses and when the film thickness is small (curves 1,1$'$,2,2$'$)
the value of $a_{opt}$ is decreased monotonically as the value of $\delta$ increases. At the same time at large values
of film thickness (curves 3,3$'$) and at high values of the losses the optimal modulation depth is increased as the
value of $\delta$ increases. Besides, as indicated in Fig.6a, with an increase of the film thickness, the value of
$a_{opt}$ for the LF resonance also increases over the whole range of losses. Nevertheless, at HF resonance (see
Fig.6b) as the film thickness increases, the value of $a_{opt}$ increases only in the case of high values of losses.
However at small values of $\delta$ the value of $a_{opt}$ for HF resonance is decreased with increasing $\delta$. It
should be noted that as it follows from the comparison of Fig.6a and Fig.6b, the value of the optimal modulation depth
for the LF resonance is less sensitive to the value of losses than the value of the optimal modulation depth for the HF
resonance.

\textbf{The dependence of the transmission resonance width upon
the film parameters} is shown in Fig.7 and Fig.8. Fig.7 shows the
dependence of transmission resonance width upon the modulation
depth. As the modulation depth increases (see Fig.7), the
transmission resonance width for small values of $a$ increases
likewise. However, when the modulation depth exceeds a certain
value, the transmission resonance width decreases with an increase
of the modulation depth. It should be emphasized that the
comparison between curves 1 and 2 (Fig.7a) shows that in the case
of LF resonance the transmission resonance width is decreased as
the film thickness increases. At the same time the comparison
between curves 1$'$ and 2$'$ (Fig.7b) results in the fact that at
HF resonance the transmission resonance width increases with
increasing film thickness. The detailed dependence of resonance
curve width upon the film thickness is presented in Fig.8. From
Fig.8 one can see the monotonous decrease of the transmission
resonance width with increasing film thickness for the LF
resonance (curve 1) and also the monotonous increase of
transmission resonance width with an increase in film thickness
for the HF resonance (curve 2).

\section{Conclusion}

In conclusion, we calculated the transmittance spectra of flat
metallic film, whose dielectric constant is periodically modulated
in one dimension parallel to the projection of the electric field
onto the film surface. The transmittancy spectra has been
calculated for the normally incident electromagnetic wave with the
wavelength close to the period of modulation. It is found that
transmittancy spectra of modulated film exhibits two maxima, the
transmittancy of modulated film in the vicinity of maxima being
much higher than that of unmodulated film. It is shown that this
abnormally high transmittancy arises due to the transformation of
incident electromagnetic wave into the symmetric or antisymmetric
SPP mode on one boundary and the subsequent transformation of the
above-mentioned symmetric or antisymmetric SPP mode into bulk
electromagnetic wave on the other boundary. This property enables
one to make such film transparent for the incident radiation and
it is shown that for small but finite absorption the transmittancy
can achieve the values of the order of unity. It is found that
when losses in the film are taken into account there exists an
``optimal tuning" of the film, i.e. the ratio between the losses
and the modulation depth of the film, which corresponds to the
highest resonance transmittancy.

The phenomenon of abnormally high resonance transmittancy of the
film at certain frequencies can be used in different devices (like
optical filters) which have to select assigned frequency.
Moreover, this phenomenon can be widely used in photo-lithography,
near-field microscopy and photonics.

\section*{Acknowledgments}

We greatly appreciate our colleague A. Nikitin for helpful
discussions.

\pagebreak .\vspace{7cm}
\begin{figure}[h!]
   \begin{center}
   \begin{tabular}{c}
   \includegraphics[260,403][501,549]{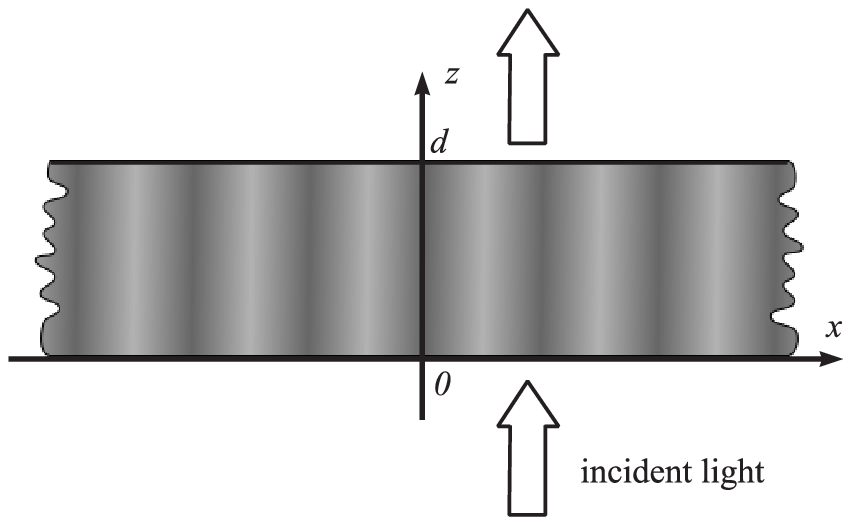}
   \end{tabular}
   \end{center}
\caption{Geometry of the problem: a metallic film of width $d$ with a modulated along x-direction dielectric
permittivity.}
\end{figure}

\pagebreak .\vspace{7cm}
\begin{figure}[h!]
   \begin{center}
   \begin{tabular}{c}
   \includegraphics[301,139][533,322]{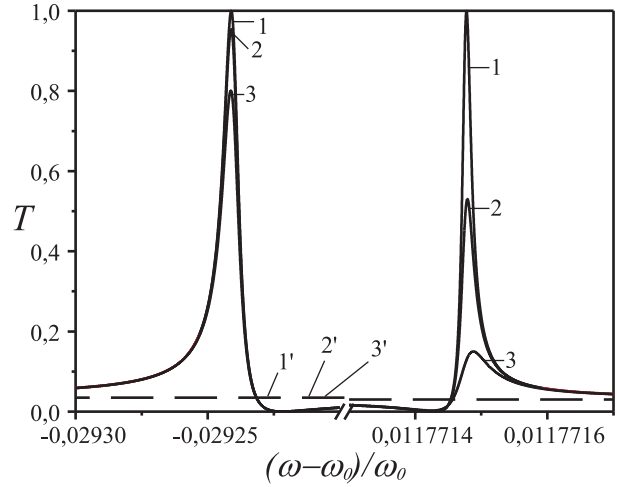}
   \end{tabular}
   \end{center}
\caption{A frequency dependence of modulated film transmittancy (solid curves 1--3) and uniform film transmittancy
(dashed curves 1$'$--3$'$) for parameters $\varepsilon_m^{\prime}=-25.0$, $\omega_0d/c=0.3$, $a=0.02$ and for three
values of losses in the film: $\delta=0$ (curves 1,1$'$), $\delta=1.0\times10^{-6}$ (curves 2,2$'$) and
$\delta=5.0\times10^{-6}$ (curves 3,3$'$). The y-axis gives the film transmittancy $T=|H_t|^2/|H_i|^2$ and x-axis gives
the dimensionless frequency $(\omega-\omega_0)/\omega_0$ ($\omega_0=qc(\varepsilon_m+1)^{1/2}(\varepsilon_m)^{-1/2}$ is
the resonance frequency of SPP at the boundary between semi-infinite metal and semi-infinite vacuum). Notice, that in
the scale of Fig.2 the difference between curves 1$'$,2$'$,3$'$ is indistinguishable.}
\end{figure}

\pagebreak.\vspace{7cm}
\begin{figure}[h!]
   \begin{center}
   \begin{tabular}{c}
   \includegraphics[253,213][517,394]{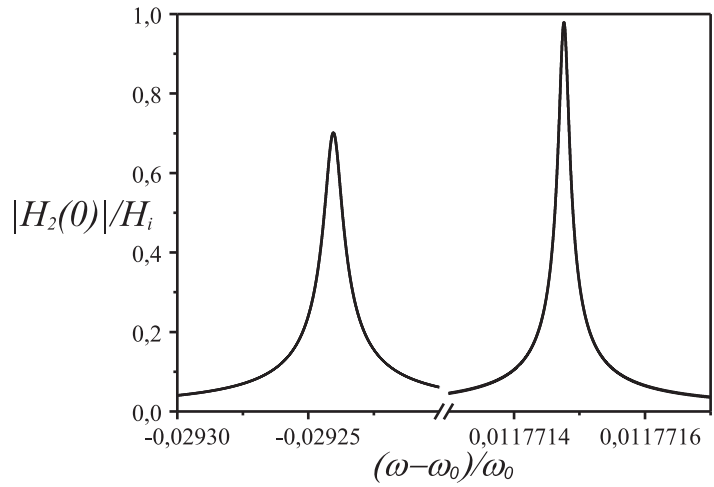}
   \end{tabular}
   \end{center}
\caption{A frequency dependence of a magnetic field spatial harmonic amplitude $H_2(0)$ in the case where
$\varepsilon_m^{\prime}=-25.0$, $\omega_0d/c=0.3$, $a=0.02$, $\delta=0$. The y-axis gives the dimensionless amplitude
of second harmonic $H_2(0)/H_i$ and x-axis gives the dimensionless frequency $(\omega-\omega_0)/\omega_0$.}
\end{figure}

\pagebreak.\vspace{7cm}
\begin{figure}[h!]
   \begin{center}
   \begin{tabular}{c}
   \includegraphics[0,0][468,142]{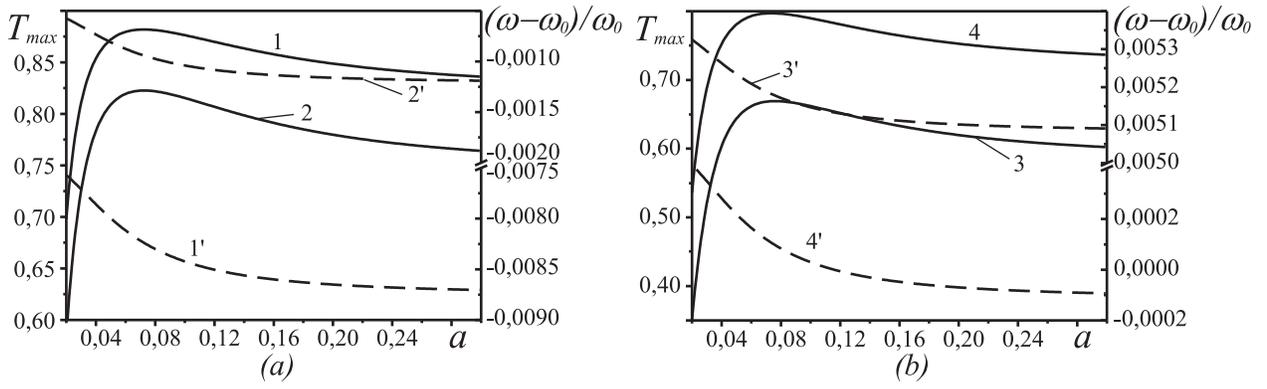}
   \end{tabular}
   \end{center}
\caption{A dependence of a transmittance maximum value $T_{max}$ (solid curves 1--4) and a resonance peak frequency
(dashed curves 1$'$--4$'$) upon a modulation depth for the LF resonance (curves 1,2,1$'$,2$'$ in Fig.4a) and for a HF
one (curves 3,4,3$'$,4$'$ in Fig.4b) for parameters $\varepsilon_m^{\prime}=-25.0$, $\delta=5.0\times10^{-6}$ and for
two film thickness values $d$: $\omega_0d/c=0.5$ (curves 1,3,1$'$,3$'$) and $\omega_0d/c=1.0$ (curves 2,4,2$'$,4$'$).
The x-axis gives the modulation depth $a$ while the left-hand side y-axis gives the maximum transmittancy at resonance
$T_{max}$ and right-hand side y-axis gives the dimensionless frequency of resonance peak $(\omega-\omega_0)/\omega_0$.}
\end{figure}

\pagebreak.\vspace{7cm}
\begin{figure}[h!]
   \begin{center}
   \begin{tabular}{c}
   \includegraphics[0,0][269,150]{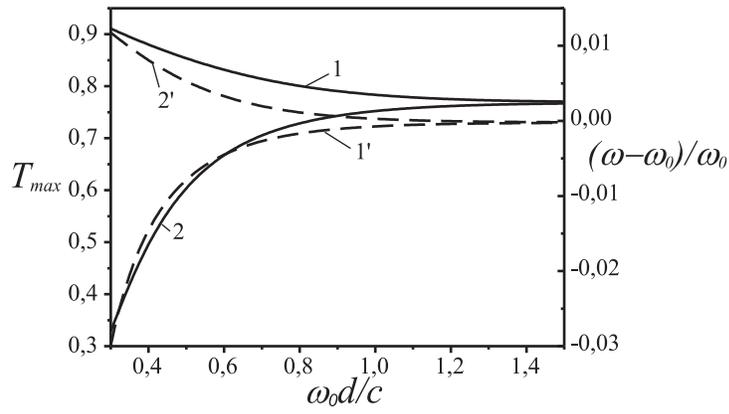}
   \end{tabular}
   \end{center}
\caption{A dependence of a transmittance maximum value $T_{max}$ (solid curves 1,2) and a resonance peak frequency
(dashed curves 1$'$,2$'$) upon film thickness for the LF resonance (curves 1,1$'$) and for a HF one (curves 2,2$'$) for
parameters $\varepsilon_m^{\prime}=-25.0$, $\delta=5.0\times10^{-6}$, $a=0.04$. The x-axis gives the dimensionless film
thickness $\omega_0d/c$, the left-hand side y-axis gives the value of transmittance maximum $T_{max}$ and the
right-hand side y-axis gives the dimensionless frequency of resonance peak $(\omega-\omega_0)/\omega_0$.}
\end{figure}

\pagebreak.\vspace{7cm}
\begin{figure}[h!]
   \begin{center}
   \begin{tabular}{c}
   \includegraphics[0,0][473,172]{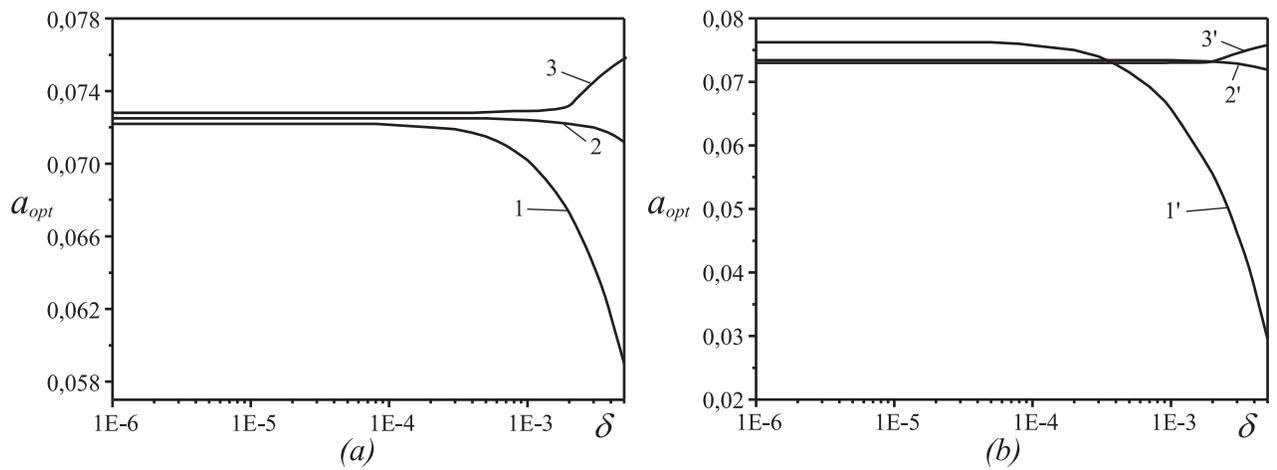}
   \end{tabular}
   \end{center}
\caption{A dependence of optimal modulation depth upon a value of losses for a LF resonance (Fig.6a) and for a HF
resonance (Fig.6b) in the case where $\varepsilon_m^{\prime}=-25.0$ and for three values of film thickness $d$:
$\omega_0d/c=0.5$ (curves 1,1$'$), $\omega_0d/c=1.0$ (curves 2,2$'$), $\omega_0d/c=1.5$ (curves 3,3$'$). The y-axis
gives the optimal modulation depth $a_{opt}$ while x-axis gives the losses in the logarithmic scale.}
\end{figure}

\pagebreak.\vspace{7cm}
\begin{figure}[h!]
   \begin{center}
   \begin{tabular}{c}
   \includegraphics[0,0][473,155]{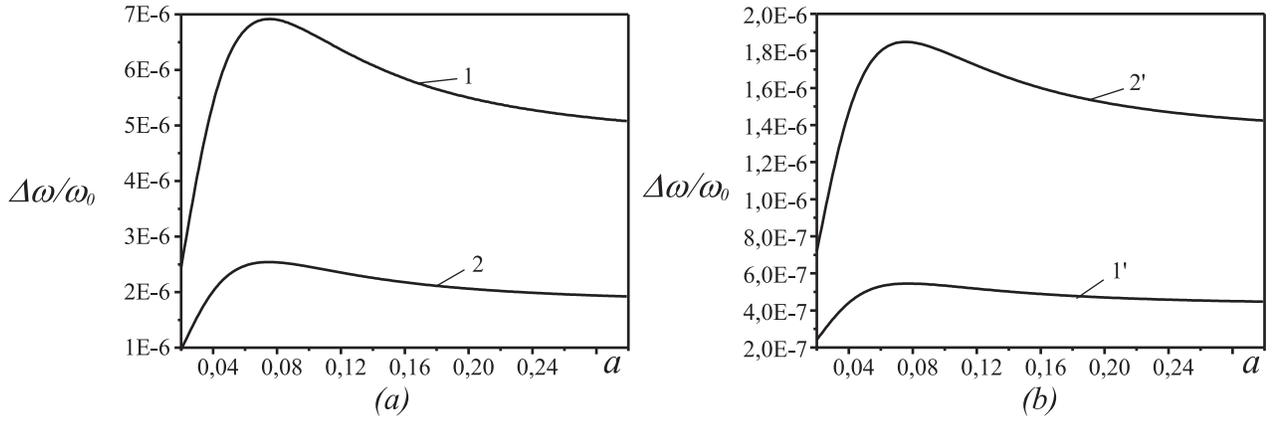}
   \end{tabular}
   \end{center}
\caption{A dependence of the transmission resonance width upon the modulation depth for a LF resonance (Fig.7a) and for
a HF one (Fig.7b) in the case where $\varepsilon_m^{\prime}=-25.0$, $\delta=5.0\times10^{-6}$ and for two values of
film thickness $d$: $\omega_0d/c=0.5$ (curves 1,1$'$), $\omega_0d/c=1.0$ (curves 2,2$'$). The y-axis gives the
dimensionless transmission resonance width $\Delta\omega/\omega_0$ ($\Delta\omega$ is the transmission resonance width
at the level of half-value of transmittance maximum) and the x-axis gives the modulation depth $a$.}
\end{figure}

\pagebreak.\vspace{7cm}
\begin{figure}[h!]
   \begin{center}
   \begin{tabular}{c}
   \includegraphics[0,0][249,160]{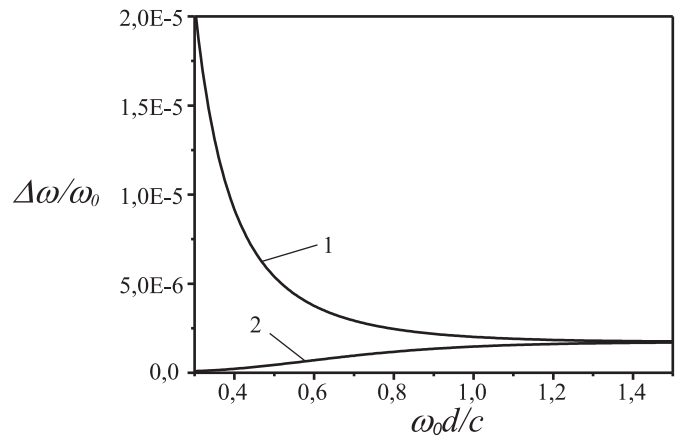}
   \end{tabular}
   \end{center}
\caption{A dependence of transmission resonance width upon film thickness for a LF (curve 1) and HF (curve 2)
resonances in the case where $\varepsilon_m^{\prime}=-25.0$, $\delta=5.0\times10^{-6}$, $a=0.04$. The y-axis gives the
dimensionless transmission resonance width $\Delta\omega/\omega_0$ and the x-axis gives the dimensionless film
thickness $\omega_0d/c$.}
\end{figure}

\end{document}